A new theory of light and colors
*E88 -- Nova theoria lucis et colorum*

Originally published in *Opuscula varii argumenti* 1, 1746, pp. 169-244
Also published in *Opera Omnia*: Series 3, Volume 5, pp. 1 - 45

Translated and Annotated
by
Sylvio R. Bistafa[*]
October 2021

Chapter II
*De formatione ac propagatione pulſuum*
On the formation and propagation of pulses in ether[a]

Foreword

In Chapter II of E88, Euler begins by rejecting Descartes' hypothesis that the universe space through which the beams of light are propagated are made of perfectly hard fixed globules, filled with his second element, and, instead, adopted the ether theory. By considering that the propagation of light in ether is similar to the propagation of sound in air, he develops a theory on the formation and propagation of pulses in ether. The pulse is supposed to be generated by a vibrating body, which propagates as a sound like wave. The position of the pulse along the propagation path is described by a cosine function whose argument resembles D'Alembert's solution of the wave equation. Based on the elastic properties of the air, he then tried to find the elastic properties of the ether, but, obviously, with no avail. Nonetheless, he correctly distinguished the local motion of the fluid particles from the motion of the pulse itself, obtaining for the speed of light in ether the value of 700,276,500 ft/s, nonetheless, the acceptable value for the speed of light is $983,571,056 \; ft/s$.

§XXV

Descartes had determined that the whole space of the universe, through which the beams of light are propagated, had fixed globules filled with his second element, having made them perfectly hard. Then, since it is clear from experience that if a series of globules of this kind is struck at one end, after a certain time, the impact is transferred to the last globule; and he thought that the rays of light were formed in a similar manner. Indeed, the Sun and other bright bodies of this kind set a constant motion on their parts, by which the globules of the second element are continually impelled, transferring the sudden pushes over long distances. Then, in fact, the light is not yet detected instantly, but only after a given time interval, in a similar manner in which the sound is propagated.

§ XXVI
If this phenomenon was known to Descartes, then, perhaps, the globules of his second element were not close to each other, but set at the smallest distances from each other, so that each one is struck before the next, it ought to be promoted through a certain small distance and, thus, it is not difficult to

---

[*] Corresponding address: sbistafa@usp.br

[a] From Middle English ēther ("the *caelum aetherum* of ancient cosmology in which the planets orbit; a shining, fluid substance described as a form of air or fire; air"). From the *Wiktionary*.

explain the succession of the spread of light. However, this explanation suffers from another extreme difficulty, which is by no means admissible. Indeed, for this propagation to occur in straight lines, it is necessary that the centers of the globules, through which the impulse is communicated, be placed in a straight line. But, this is incompatible with the principles of geometry, because when many globules are arranged in such a way, their centers are arranged in straight lines according to each direction.

§ XXVII

Having, therefore, repudiated this explanation, we will safe move on in this investigation, if we conceive that the propagation of light is similar to the propagation of the sound. However, the propagation of the strongest sound through the air, that is an elastic fluid, which is not only endowed with immense force of expansion, but can also be compressed by any impulse to a greater degree of condensation. Therefore, it will be advisable to conceive the same character to the ether, so that through it, the light can be thought in a similar manner in which the sound is propagated through the air. Wherefore, the explanation of the rays of light will depend on the investigation of the nature of elastic fluids, which is a most difficult matter to explore in mechanics, I will endeavor to explain how a pulse is produced and propagated through such fluid, and in whatever manner may be understood.

§ XXVIII

Since we know from nature of sound that the pulse in the air is excited by the vibratory motion of a body, we rightly conclude that the pulse is generated in a similar manner in the ether and in all other elastic media. Therefore, we may be able to understand more easily how to represent the formation of this kind of pulses, and how they are transferred further through the elastic medium, by considering the body $EAF$ (Fig. 1), fixed at points $E$ and $F$ and distributed along $A$ like a vibrating cord, in this state, all the particles in a line $AO$ possess the natural density and elasticity of the elastic medium. As soon as the cord is released from $A$, it assumes the shape $EaF$, where the adjacent particle at the middle point $A$ is displaced as far as $a$, and the particles of the medium beyond $a$ suffer the greatest condensation; for that reason, the particles which were distributed along the line $AO$ are now driven to a smaller space $aO$.

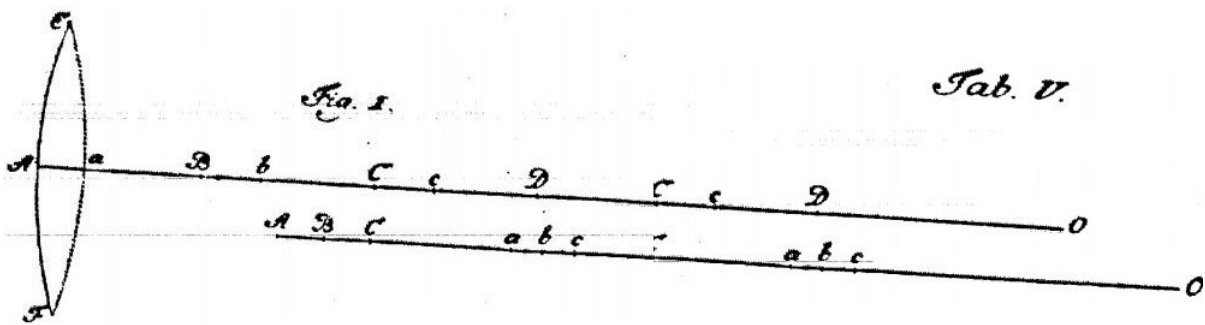

Figure 1: A cord vibrating in ether. (Adapted from Euler, 1746, Table V, Fig. 1)

§ XXIX

It is clear that when the cord assumes the shape $EaF$, the particle $a$ of the elastic medium is in the state of maximum condensation, and, therefore, it will achieve the greater expected elasticity than the remaining particles situated in the direction of $O$. Hence, the condensation of the particles beyond $a$ will decrease continuously, and yet, on account of a non infinite elasticity, in the mean time, the condensation cannot extend to infinity. Therefore, let us consider that the effect of the vibration extends itself as far as $B$, and since the condensation of the particle $a$ is maximum, the condensation of the sequence of particles decreases continuously, such that in $B$, the density will be equal to the natural density: meanwhile beyond $B$, and as far as $O$, all the matter of the elastic fluid will remain in the natural state of density and elasticity.

§ XXX

Now, whether the chord in the position $EaF$ stays or recedes, the particle $a$, because of the resulting increased elasticity, will expand itself, and by impelling the following particles, they will condensate even more. In this way, the density in $a$ decreases continuously, until the natural state is restored: however, once the particle $B$ is driven, it will not only be compressed, but also it will advance by a certain distance $Bb$, such that now, the maximum compression is found in $b$. However, meanwhile, the particle $B$ is now transported to $b$, and each of the following particles will condensate even more beyond the natural state, and this change will extend itself as far as $C$. Afterwards, in a similar way, now the particle in $b$ will relax itself, and the point $C$ further away will be pushed to $c$, and from here, the condensation pulse will be propagated to the point $D$ further away; and in this way, the impulse originated in $A$, will be continuously propagated; and, eventually, it will reach the farthest limit $O$.

§ XXXI

Therefore, it is clear that the state of maximum compression which after the cord had been put into vibration appears in $a$, it then successively propagates further away through the line $AO$, so that it is transferred to each intermediate point all the way to $O$. During the time that this pulse is transmitted from $a$ to $O$, the series of particles of the elastic media will not be in equilibrium, but, in some places a particle will be more compressed than the others, and therefore more elastic. However, the adjacent particles on both sides will be less compressed, and this inequality of compression will not extend everywhere beyond a given distance. So that if we consider that the maximum compression is at point $b$, the following particles up to $C$ will be justifiably more compressed than those situated beyond $C$, with these particles remaining in their the natural state: in a similar way, also the matter situated backwards from $b$, up to a given distance, where the natural state will be found, on the supposition that there will be no new impulse of the cord.

§ XXXII

Therefore, as long as a single pulse originated at the interval $Aa$ propagates through the distance $AO$, the subtle matter will not be simultaneously agitated everywhere, but with a continuous propagation of the impulse, in a given local magnitude, outside of which the elastic matter is both in equilibrium and in the natural state. This is the place where the state of equilibrium is destroyed, which is usually called a pulse, and, therefore, each pulse continuously propagates further away, and it proceeds like a wave on the surface of the water, so that the material itself is not simultaneously transported along, but only the state of greater compression moves through successive parts. In the mean time, however, each particle, while the pulse clings in that place, moves forward a little and step back; whose motion is rightly distinguished from the motion of the pulse itself.

§ XXXIII

The pulse, therefore, at any moment in line $AO$ will occupy a certain distance, in which the elastic particles are placed in a state other than the natural one, so that the particles of the medium outside this space, maintain their natural density and elasticity. Therefore, the particles of the medium in the very place of the pulse will be in continuous agitation; for while some particles relax themselves, others particles become more condensed and move forward a little; hence, the pulse itself will be transferred further. Accordingly, this explanation will consist of two parts, one of these, is the propagation of the pulse, the other, particles to which the pulse is bounded. It could happen that both motions may be extremely disturbed by the agitation of the particles, which in our considerations will be considered minimal, the motion as a whole will soon result in a certain uniform law; the uniformity of which nature constantly strives in all the smallest motions.

§ XXXIV

Not only does it seem to be consistent with the truth that the pulse progression is uniform, and experience also teaches us that sound propagates through the air with a constant motion. Therefore, since it is extremely difficult to determine the motion of the pulses as well as to determine a priori the motion of the particles from mechanical principles, our investigation will lighten up if we assume like Newton, that the motion of the pulse is uniform, and that each particle in the pulse motion is similar to the motion of a pendulum completing the smallest of free oscillations. Then, indeed, the mechanical principles will not only allow us to show how a motion of this kind can subsist, but also its magnitude and the true velocity of the pulses can be assigned.

§ XXXV

Therefore, the method that we will use to expedite this task will be entirely unique, partly supported on conjecture and partly on established mechanical principles. For by conjecture, which experience indeed confirms, we now assume, as it were known, what will be the motion of the pulse, and then the internal motion of the particles. The theory, together with this conjecture, will first show that such motion exists in an elastic fluid: moreover, the density and the elasticity of the subtle matter itself will reveal the shape and magnitude of this motion in general. The great Newton first pursue the same path in the development of Proposition 47, Book II, *Principia*, where he also defined the velocity of motion of the pulse, and from here, he derived the propagation of sound; and by this technique, I have carried out another most difficult investigation to the end.

§ XXXVI

First of all, to follow the footsteps of this Most Intelligent Man, the most difficult task is the analysis considered in Proposition XLVII, Book II, which, to most people, is not a little obscure; I will elucidate it, and, henceforth, I will thereafter adapt it to the present investigation. Let us consider $AB$ and $BC$ as two small particles (Fig. 2) of an elastic fluid located in the straight line $AO$, which still dwell in the natural state. Set $AB = BC = c$, $D$ the density of the fluid in the natural state, and $E$ the elastic force. Moreover, let us equate this elastic force to the weight of a cylinder of height $= k$, which is filled with a equally dense heavy matter, being this an elastic fluid. Even though, indeed, ether lacks gravity, yet, it is possible to conceive a heavy matter, whose density is equal to the density of the ether: then, the elastic force $E = Dk$, where $Dk$ is the weight of the cylinder with a height $= k$ and density $= D$.

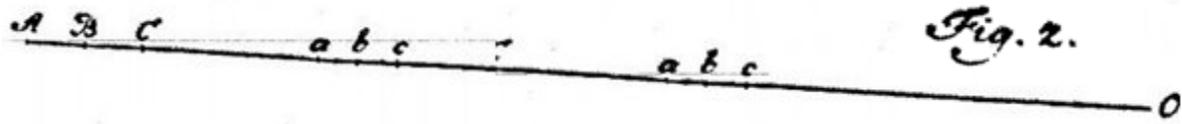

Figure 2: Two small particles $ab$ and $bc$ of an elastic fluid located in the straight line $AO$. (Adapted from Euler, 1746, Table V, Fig. 2)

§ XXXVII
Let us assume that at the present instant, the pulse has already arrived at point $B$, and that this point is now the first to begin to be excited by the motion. Suppose the pulse is propagated through the space $AO = a$ in a time $T$; and, therefore, the pulse will reach point $C$ after the time $\frac{c}{a}T$ has elapsed. But after the pulse reached the point $B$ at time $t$, we assume that the point $B$ was transferred by the wave to point $b$; since the motion of this point is similar to the motion of a pendulum performing the smallest oscillations; the space $Bb$ will be proportional to the *sine verso*[b] of a certain angle, which is proportional to the time $t$, set this angle $= mt$, such that $Bb = \propto versin(mt) = \propto (1 - \cos mt)$.

§ XXXVIII
But since each of the particles, after they began to be induced to move, are moved in a similar manner, the point $A$, which has already endured the impulse for a period of time $t + \frac{c}{a}T$, will now be found in $a$, then, $Aa = \propto \left[1 - \cos m\left(t + \frac{c}{a}T\right)\right]$. But point $C$, whose motion lasted through the time $t + \frac{c}{a}T$, will be transferred to $c$, that is, $Cc = \propto \left[1 - \cos m\left(t - \frac{c}{a}T\right)\right]$. Henceforth $ab = Bb + AB - Aa$, and $bc = Cc + BC - Bb$, and thus:

$$ab = c + \propto \cos m\left(t + \frac{c}{a}T\right) - \propto \cos mt$$
$$bc = c + \propto \cos mt - \propto \cos m\left(t + \frac{c}{a}T\right)$$

or

$$ab = c - \propto \cos mt + \propto \cos mt \cdot \cos \frac{mc}{a}T - \propto \sin mt \cdot \sin \frac{mc}{a}T$$
$$bc = c + \propto \cos mt - \propto \cos mt \cdot \cos \frac{mc}{a}T - \propto \sin mt \cdot \sin \frac{mc}{a}T$$

§ XXXIX
Since the motions of the particles sustained by the pulse are always small, the intervals $ab$ and $bc$ will differ as little as possible from the natural magnitude $AB = BC = c$, hence the quantity $\propto$ will be very small with respect to $c$. And, since the interval $c$ is also small, then $\cos \frac{mc}{a}T = 1 - \frac{m^2c^2}{2a^2}T^2$ and $\sin \frac{mc}{a}T = \frac{mc}{a}T$.
Whence, we set

$$ab = c - \frac{\propto mc}{a}T \sin mt - \frac{\propto m^2c^2}{2a^2}T^2 \cos mt,$$
$$bc = c - \frac{\propto mc}{a}T \sin mt + \frac{\propto m^2c^2}{2a^2}T^2 \cos mt.$$

---

[b] The *sine verso* is a trigonometric function little used today. It is usually written as *versin* or *vers* and is defined as: $versin(\theta) = (1 - \cos\theta)$

Hence, the density of the particle $ab$ will be $= \frac{AB}{ab}D = \frac{Dc}{ab}$ and the elastic force, which for the smallest variations follows the ratio of densities will be $= \frac{Ec}{ab} = \frac{Dck}{ab}$. Similarly, the density of the particle $bc$ will be $= \frac{Dc}{bc}$, and the elastic force will be $= \frac{Dck}{bc}$. Hence, the excess of the elastic force in $ab$ over the elastic force in $bc$ will be $= \frac{Dck(bc-ab)}{ab \cdot bc} = \frac{Dck}{ab \cdot bc} \cdot \frac{\propto m^2 c^2}{a^2} T^2 \cos mt$.

§ XL

But since the intervals $ab$ and $bc$ little differ from their natural magnitude $c$, it is safe to write in the present formula $ab \cdot bc$ as $c^2$. When this is done, the excess of the elastic force in $ab$ over the elastic force in $bc$ will be $= \frac{\propto m^2}{a^2} Dck\, T^2 \cos mt$, and this will be the force by which the point $b$ is moved further away towards $O$. Moreover, to find the velocity by which the point $B$ advances towards $O$, it may be required to find the differential of the distance $Bb = \propto (1 - \cos mt)$ which is $= \propto m dt \sin mt$, which, when divided by the element of time $dt$ will give the velocity $= \propto m \sin mt$, which is due to the height $v$; such that $\sqrt{v} = \propto m \sin mt$, and $v = \propto^2 m^2 \sin^2 mt$.

§ XLI

Have been found the velocity of point $b$, which will be common to all the points of the particle $ab$, then, the space increment can be found from the driving force $\frac{\propto m^2}{a^2} Dck\, T^2 \cos mt$, while point $b$ advances through the small distance $\propto m dt \sin mt$. Since, indeed, the mass of the particle $ab$ is $= Dc$, then, from the principles of mechanics $Dc \cdot dv = \frac{\propto m^2}{a^2} Dck\, T^2 \cos mt \cdot \propto m\, dt \sin mt$. But, $dv = 2 \propto^2 m^3 dt \sin mt \cos mt$, which when substituted in the latter expression, will give the following equation $2 \propto^2 m^3 Dc\, dt \sin mt \cos mt = \frac{\propto^2 m^3}{a^2} Dck\, T^2 dt \sin mt \cdot \cos mt$; whence $2 = \frac{k\, T^2}{a^2}$, and $\frac{a}{T} = \sqrt{\frac{k}{2}}$, where it is seen that $\frac{a}{T}$ expresses the velocity at which the pulse is propagated, which, therefore, is so great as that which is generated from a heavy body falling from the height $\sqrt{\frac{k}{2}}$.[c]

§ XLII

Therefore, the velocity of the pulse is determined from the density and the elasticity of the medium; and, thence, it is possible to define how much time it takes to traverse a given distance. But the actual motion of the individual particles in the pulse is not determined, because the letters $a$ and $m$ have disappeared from the calculation, and, thus, they remain indeterminate. Naturally, the agitation of the particles will depend everywhere in the first impressed motion by the pulse in the elastic fluid, by a cord or whatever means, it will be kept forever similar to it. Whence, it is understood that by whatever manner the fluid was struck at the beginning, by a cord or other means, whether stronger or weaker, the pulse, however, is nevertheless propagated with the same velocity, since it depends only on the letter $k$ or from the elasticity applied to the density. This is also constantly asserted from experience, whereby the most powerful sounds, as those of the artilleries, which are observed to be propagated with the same velocity, as the most weak sounds.

---

[c] According to Euler, the $Speed\ of\ Sound\ c$ is the velocity that a body acquires when falling from a height $= \sqrt{\frac{k}{2}}$. Hence, $c = \sqrt{gk}$, where $g$ is the gravity. However, it is known that $c = \sqrt{\frac{B}{D}}$, where $B$ is the *Bulk Modulus* (or *Elasticity*) which is the reciprocal of *Compressibility*. The air *Isothermal Bulk Modulus* is equal to the *Applied Pressure p*, and, then, $c = \sqrt{\frac{p}{D}}$. The pressure $p$ at the base of a cylinder with a height $= k$ and filled with air with density $= D$ is $p = Dgk$, giving $c = \sqrt{gk}$, which agrees with Euler's result for the speed of sound. Also, since $gk = \frac{p}{D}$, and $E = Dk$, then, the elastic force $E = \frac{p}{g}$.

§ XLIII

Hence, since both the density and elasticity of the air are known, the speed of the pulse excited in the air and the speed with which the sound is propagated can be ascertained. If, indeed, the elasticity of the air is equal to the weight of a column of mercury 30 inches high because mercury, and because the ratio of the specific weight of the mercury to that of the water is as 13593 to 1000, and water to air, according to Newton, is as 870 to 1, the elastic force of the air will be equal to the weight of a column of air whose height is $= 30 \cdot 13.593 \cdot 870$ British inches, that is 29565 feet or 28678 Rhenish feet. Here we will have that $k = 28678\ Rhenish\ feet$ and $\frac{1}{2}k = 14332\ Feet$. However, a body descending from such height acquires a velocity which in one second traverses a distance of $947\ Rh.ft.$ or $975\ Brit.ft.$ However, Newton assumed that the mercury was related to water as 13 to 1, and found a space of 979 feet to be completed in one second.

§ XLIV

But although experience can reveal a greater speed of sound, since, indeed, it is agreed that sound covers a space of about 1100 feet in one second[d], however, whatever is the cause of this difference, there is no doubt that the speeds of pulses propagating through different elastic media are bound to the square root of the height $k$. Or, since, in any elastic media, $k = \frac{E}{D}$, the velocity of the pulses is composed of the square root of the ratio of elasticity to the density of the medium. Furthermore, in the above expression for the elastic force $\frac{Dck(bc-ab)}{ab \cdot bc}$, had we put $c^2$ in place of $ab \cdot bc$, would result in a slightly greater value, the elastic force that we assume rather small, which if a little greater than it is determined, will also produce a greater speed of the pulse; whence it is evident that the theory itself conspires with experience, and the conclusion thence, truly slightly reduced.

§ XLV

Although the method which is here employed after Newton is indirect, and to a perfect theory of propagation of pulses in an elastic fluid is still far distant, yet, the pulse velocities which are excited in different elastic media may be rightly compared to each other: since the velocity of the particles does not depend on their motion, for its determination the mechanical principles are still, perhaps, not sufficiently known. Therefore, since the velocity of sound or of the pulse in air is known, likewise, it will be possible to be determined for any elastic medium for which both density and elasticity with respect to the air can be known. Therefore, if we assume that in ether the ratio is $m$ and $n$ times greater than in the elastic air, then the velocity at which the pulse is propagated in the air will be to the velocity at which the pulse is propagated in the ether as 1 to $\sqrt{mn}$.

---

[d] Sir Isaac Newton's 1687 Principia includes a computation of the speed of sound in air as 979 feet per second (298 m/s). This is too low by about 15%. The discrepancy is due primarily to neglecting the (then unknown) effect of rapidly-fluctuating temperature in a sound wave (in modern terms, sound wave compression and expansion of air is an adiabatic process, not an isothermal process). At 20 °C (68 °F), the speed of sound in air is about 343 meters per second (1,235 km/h; 1,125 ft/s; 767 mph; 667 kn), or a kilometer in 2.9 s or a mile in 4.7 s.

§ XLVI

Therefore, if the values of the letters $m$ and $n$ are known, the speed of light or the pulse excited in the ether can be determined from these values. Since neither the density nor the elasticity of the ether can be determined from experiments yet, nonetheless, the speed of light is known with satisfactory accuracy, hence, also the value of the formula $\sqrt{mn}$, and, from this, the intrinsic natural characteristics of the ether will be known. The light, however, is observed to reach us from the sun at an interval of about $8'$, and, because the sound in the interval of $1''$ completes a distance of 1040 Parisian Feet, and, therefore, in the time interval of $8'$ about $500000 \; Paris \; Ft.$: the speed of sound to the speed of light will hold as the distance of $500000 \; Paris \; Ft.$ to the distance of the Earth to the Sun, and in the same ratio as 1 to $\sqrt{mn}$.

§ XLVII

Let us assume that the semi diameter of the Earth is equal to $19615791 \; Paris.Ft$, and let us set the horizontal parallax of the Sun as $13''$, giving the distance of the Earth to the Sun as equal to 15866 times the semi diameter of the Earth, resulting in $311234300000 \; ft.$ [e] Therefore, we will have that:
$$1:\sqrt{mn} = 5:3112343$$
then,
$$\sqrt{mn} = 622468 \text{[f]}$$
hence, it will be found that
$$mn = 387467100000.$$
If, therefore, the rarefaction of the ether is known, it will be found at once how much its elasticity is greater than the elasticity of the air; and, in turn, from the elasticity of the ether, its density could be inferred. But, since no sensible reduction of motion is observed in the movement of several planets, it is necessary that the density of the ether be quite small and, possibly, more than $100000000$ times less than the density of the atmosphere.

§ XLVIII

---

[e] Solar Parallax
    $\varpi_\odot$ = 13''= $6.3 \times 10^{-5} rad$.
    $R_\oplus$ =$19615791 ft$
    $a$ = $311,234,300,000 \; ft = 94,864,215 \; km$ (The acceptable distance from Earth to the Sun is $149,597,870 \; km$. This is called the astronomical unit, or AU, which is used to measure distances throughout the solar system)

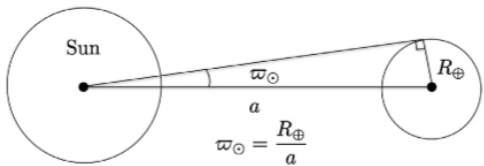

[f] Since the velocity at which the pulse is propagated in the air will be to the velocity at which the pulse is propagated in the ether as 1 to $\sqrt{mn}$, and knowing that the velocity of sound is equal to $1,125 \; ft/s$, then, the velocity of light in ether would be given by $v_{light} = \sqrt{mn} \cdot c = 622468 \cdot 1,125 \text{ft/s} = 700,276,500$ ft/s. However, the accepted value for the speed of light is $983,571,056 \; ft/s$.

In the following discussion about the relaxation of the motion of the planets arising from the resistance of the ether, the most common argument is that the motion of planets is not disturbed provided the density of the ether is 387367100 times less than that established for the air; neither would be the force inferred from observations, even if it would be determined that it is still notably greater. Hence, as far as the elasticity of the ether is concerned, the force would remain at least a thousand times greater than the elastic force of the air, and nothing would prevent it from being taken twice or even three times greater. However, the elastic force is so great that it seems to be fully capable of producing all the phenomena, which are commonly ascribed to the elastic force of the ether, to explain the hardness of bodies and their elasticity.

§ XLIX

As far as the hardness of our bodies is concerned, if the elastic force is not exactly defined from experiments, nonetheless, we will be able to assign limits to it, which surely may be surpassed. May be chosen a very hard body, and from it we fabricate a slender cylinder, whose base, for example, is set as a Rhenish square feet scruple[g] or $\frac{1}{1000000}$ $square\ feet$, and it is asked the weight $P$ of this divided prolonged cylinder when laid straight. Since the pressure of the atmosphere in this base[h] corresponds to the weight of $\frac{32}{1000000}$ $cubic\ feet$ of water, this will give a force of $\frac{2240}{1000000}$ $pounds$. Hence, since $P$ is the effect of the elastic force of the ether, then, the elastic force of the ether will be at least $\frac{100000\,P}{224}$ times greater than the elasticity of the air[i], if indeed the weight $P$ is expressed in pounds.

§ L

But, from several experiments of this kind, the elasticity of the ether has never been a thousand times greater than the elasticity of the air, but it is always much smaller. Neither, however, were the hardest bodies employed, nor the ether exerted pressure over the whole base, but only on that part which is impervious to it. Therefore, the value of the letter $n$ cannot be less than 1000: then, let us set $n = 1000$, giving $m = 387467100$ or about 400000000, which is a value also known in the investigation of the motion of planets disturbed by the resistance of the ether, and showing that it was not possible to bring forth so great an effect, that might oppose the observations.

§ LI

From these, therefore, it is understood, how the pulse is formed in the ether and propagated to the greatest distances. In the same way as in the air, if pulses are excited and propagated from a trembling body, so, in the same way, to arouse a pulse in the ether, it is necessary that it be struck somewhere by some force, and there, the perturbation of the state of equilibrium, no doubt, will bring about a pulse that will propagate itself further away: in the same way in every direction, by which we have shown just in one direction how it should happen. And, likewise, the pulse is propagated to the eye in a similar way, and by exciting the nerves, it will arouse the sense of vision. Moreover, although for this succession it is required the most frequent sequence of pulses, however, the effect of each of all the effects must be judged.

---

[g] A scruple is equal to a thousandth part of Rhenish feet.

[h] The atmospheric pressure is equal to 14.7 psi or 2116.8 lb/ft², resulting in a force of approximately 0.002 pounds in a $\frac{1}{1000000}$ ft² base. By its turn, the specific weigh of the water is equal to 62.43 pounds-force per cubic foot, then, $\frac{32}{1000000}$ ft³ weights approximately 0.002 pounds, whereas Euler gives $\frac{2240}{1000000} = 0.002240$ pounds.

[i] $\frac{F_{elastic\ air}}{F_{elastic\ ether}} = \frac{224}{100000\,P} \Longrightarrow F_{elastic\ ether} = \frac{100000\,P}{224} \cdot F_{elastic\ air}$.

§ LII

Then, we also saw that the pulse, once it was formed, moved in a straight line, since the medium was considered uniform; whence, at the same time, it is realized the straightness of the rays of light. The propagation of the pulse itself arises from the agitation of the particles of the elastic medium where the pulse is engaged, and since everywhere it turns toward a definite direction, it induces the same motion in the direction of the pulse. Hence, therefore, every difficulty recalled above by Newton does not disappear, as by this theory the light that had been admitted through a hole in the dark chamber ought to be diffused in every direction. The reasoning why the pulse deflects laterally and also returns backwards stands opposite to both theory and experience. And so this is clearly explained from the theory of Newton itself, whence, all the more surprising, this propagation of pulses was rejected as being inappropriate in the explanation of light.